\documentclass[preprint,preprintnumbers,showpacs,amsmath,amssymb]{revtex4-1}
\usepackage{graphicx}
\usepackage{dcolumn}
\usepackage{bm}
\usepackage{bbm}

\begin{document}

\title{Quantum Parameter Estimation in the Unruh-DeWitt detector model}

\author{Xiang Hao}
\affiliation{School of Mathematics and Physics, Suzhou University of Science and Technology,
\\Suzhou, Jiangsu 215011, People's Republic of China\\}

\affiliation{Pacific Institute of Theoretical Physics, Department of Physics and Astronomy,
\\University of British Columbia, 6224 Agriculture Rd., Vancouver B.C., Canada V6T 1Z1.}

\email{xhao@phas.ubc.ca}

\author{Yinzhong Wu}
\affiliation{School of Mathematics and Physics, Suzhou University of Science and Technology,
\\Suzhou, Jiangsu 215011, People's Republic of China\\}

\begin{abstract}

Relativistic effects on the precision of quantum metrology for particle detectors, such as two-level atoms are studied. The quantum Fisher information is used to estimate the phase sensitivity of atoms in non-inertial motions or in gravitational fields. The Unruh-DeWitt model is applicable to the investigation of the dynamics of a uniformly accelerated atom weakly coupled to a massless scalar vacuum field. When a measuring device is in the same relativistic motion as the atom, the dynamical behavior of quantum Fisher information as a function of Rindler proper time are obtained. It is found out that monotonic decrease in phase sensitivity is characteristic of dynamics of relativistic quantum estimation. To improve relativistic quantum metrology, we reasonably take into account two reflecting plane boundaries perpendicular to each other. 

\vspace{1.6cm} Keywords: phase quantum estimation, relativistic effects, Unruh-DeWitt model, quantum Fisher information

\end{abstract}

\maketitle

\section{Introduction}

The combination of relativity and quantum theory contributes to a fruitful area of research, which in particular includes relativistic quantum information theory and technology \cite{Peres04,Friis12,Bruschi13,Martin13,Martin14,Landulfo09,Celeri10,Wang14,Salton15,Su14}. In information processing, particle detectors or observers moving along relativistic trajectories, can send and receive signals which are relevant to internal degrees of freedom of detectors. The role of detectors is to locate and label events in curved spacetimes. Recently, the use of non-inertial motions to perform quantum computing has been investigated by means of accelerated optical cavities \cite{Friis12,Bruschi13} or atoms in relativistic regimes \cite{Martin13,Martin14}. The relativistic technology contributes to ultra-fast quantum computation which can overcome some obstacles from decoherence. In this Letter, we employ this idea to study the relativistic effects of accelerated qubits on the precision of parameter quantum estimation.

The selection of a physically implementable model is a key to our scenario. For many years, the Unruh-DeWitt ($\mathrm{UdW}$) model \cite{Unruh1976,Luis08,DeWitt1967,Brown13} have been extensively applied to the study of aspects of quantum field theory in curved spacetimes. This model is referred to as a particle detector, like a two-level atom, which is linearly coupled to a quantum scalar field with simple monopole interactions. In comparison of $\mathrm{QED}$, the model characterizes adequately light-matter interactions \cite{Scully1997} in some specific settings. One great success of the model lies in the demonstration of Unruh effect. It was shown that a uniformly accelerated detector behaves as an inertial detector in a thermal bath, with a characteristic temperature, i.e., Unruh temperature, proportional to its proper acceleration. This open quantum system approach is applied to quantum gravity on some spacetimes \cite{Yu11,Feng15}. The phenomenological model can be simulated in trapped ion systems or superconducting circuit experiments \cite{Nation09,Wilson11}. Our scheme is to explore quantum metrology by the $\mathrm{UdW}$ model.

As we know, quantum Fisher information ($\mathrm{QFI}$) is thought of as one efficient measure for parameter quantum estimation \cite{Braunstein1994, Petz1996}. Recently, $\mathrm{QFI}$ has been widely studied in various fields involving the investigation of uncertainty relations \cite{Luo00,Gibilisco07}, the estimation of quantum speedup limit time \cite{Taddei13}, the characterization of quantum phase transition \cite{Invernizzi08}, and the detection of entanglement \cite{Boixo08}. Until now, some interesting works have considered the Unruh effect on quantum metrology in the non-inertial frame \cite{Aspachs10,Yao14,Hosler13,Ahmadi14,Tian15}. The authors have found that Fork states can achieve the maximal $\mathrm{QFI}$ in the presence of scalar fields in a $1+1$-dimensional Minkowski spacetime \cite{Aspachs10}. Dirac fields in non-inertial frames were investigated in \cite{Yao14}. Additionally, relativistic effects were considered as one resource for performing quantum metrology \cite{Ahmadi14}.The Fisher information for population measurement has been evaluated by the use of the open quantum system method \cite{Tian15}. When a measuring device is in a relativistic motion, observation occurs in a non-inertial frame. Our aim is to determine some common features of relativistic quantum metrology. We try to find some controllable conditions that contribute to the enhancement of relativistic quantum metrology.

In this paper, the dynamics of $\mathrm{QFI}$ with respect to phase is investigated under the condition that a uniformly accelerated atom is weakly coupled to a massless quantum scalar field in the presence of reflecting boundaries. The paper is organized as follows. First, the $\mathrm{UdW}$ model and one calculation method of $\mathrm{QFI}$ are presented. We choose a pure state with a phase parameter as an initial state, and describe the evolution of the accelerated atom by means of open quantum system approach in the non-inertial frame. Second, in accordance with motions of a measuring device, the calculations of $\mathrm{QFI}$ as a function of the proper time in the non-inertial frame are demonstrated in detail. The effects of reflecting boundaries on phase quantum estimation are considered. Finally, a conclusion depicts key findings about relativistic parameter quantum estimation and some possible physical realizations.

\section{Model and Dynamics}

What we are interested in is the dynamical behavior of parameter quantum estimation related to a uniformly accelerated particle detector, such as a two-level atom. In this scenario, we present the Unruh-DeWitt detector model to characterize the atom coupled to a massless quantum scalar field. In the weak coupling limit, atomic transitions with no exchange of angular momentum is reasonably considered \cite{Martin14}. Without loss of generality, the Hamiltonian of the model can be expressed as
\begin{equation}
\label{eq:(1)}
H = H_0^{(d)}+H_0^{(f)}+H_{I}.
\end{equation}
The system Hamiltonian of the detector is written as $H_0^{(d)}=\frac 12 \omega_0 \sigma_z$. $\omega_0$ represents the energy gap between the ground state and excited state. $\sigma_z$ is the Pauli matrix for the detector. Thus, the quantum states of the detector can be described in terms of a two-dimension Hilbert space. The form is written as a $2\times 2$ density matrix which is a completely positive and Hermitian operator. $H_0^{(f)}$ is the standard Hamiltonian of massless scalar fields, which is determined by the massless Klein-Gordon equation. The details of the field need not be specified here. The interaction Hamiltonian $H_{I}$ between the detector and scalar field is given by
\begin{equation}
\label{eq:(2)}
H_I = \lambda \, \left[ \sigma_z \phi \big(x(\tau)\big)  \right],
\end{equation}
where $\lambda$ is the strength of the interaction. The weak coupling condition of $\lambda \ll \omega_0$ is assumed so that we can obtain a completely positive dynamical map for the detector. The field operator $\phi\big( x(\tau) \big)$ is dependent on the worldline of the accelerated detector.

For the description of the dynamics of the system, the composed system is initially prepared in a factorized state, with the detector at rest and the scalar field in the vacuum state $|0_f\rangle$. The two sets of spacetime coordinates are involved in this model. One denotes $(\tau,\xi)$ in a non-inertial frame that is moving together with the atom, and the other one is $(t,x)$ in the inertial frame. The relation between two sets of spacetime coordinates holds that
\begin{align}
\label{eq:(3)}
ct = & \xi \, \sinh(\frac {a\tau}{c}) \nonumber \\
x = & \xi \, \left[ \cosh(\frac {a\tau}{c})-1  \right]
\end{align}
where the trajectory of constant Rindler position $\xi=\frac {c^2}{a}$ is used to describe the observation of uniformly accelerated atom. The worldline is just shown by Eq(3). For simplicity, it is assumed that the natural unit $c=1$ and that the atom is in the position $x=0$ at the proper
time $\tau=t=0$. Under the circumstance of weak couplings, we use the perturbation method to obtain the dynamics of the accelerated detector. It is a natural choice to establish a dynamical map in the non-inertial frame. In the interaction picture, the dynamics of the reduced density for the detector after tracing over the field degrees of freedom is expressed as
\begin{equation}
\label{eq:(4)}
\frac {\mathrm{d}\rho(\tau)}{\mathrm{d}\tau} = -\int_{0}^{\tau} \; \mathrm{d}\tau_1\; \mathrm{Tr}_{f} \left\{ \; \left[ H_I(\tau),[H_I(\tau_1),\rho(\tau)\otimes \rho_f] \;  \right] \; \right\},
\end{equation}
where $H_I(\tau)$ is the Hamiltonian of the total system in the interaction picture. $\rho(\tau)$ represents the state of the detector in the non-inertial frame, and $\rho_f$ denotes the vacuum field in the non-inertial frame. By the approach of \cite{Benatti03, Benatti04}, the analytical solution to the dynamical equation is equivalent to the quantum master equation which has the Lindblad form of
\begin{equation}
\label{eq:(5)}
\frac {\mathrm{d}\rho(\tau)}{\mathrm{d}\tau} = -i[H_{eff}, \rho(\tau)]+L[\rho(\tau)].
\end{equation}
The key to the quantum master equation is the dissipation term given as
\begin{equation}
\label{eq:(6)}
 L[\rho(\tau)] = \frac 12 \sum_{k=\pm,z} \gamma_k \; \left( 2\sigma_k \rho \sigma^{\dag}_{k}- \sigma^{\dag}_{k}\sigma_k \rho-\rho\sigma^{\dag}_{k}\sigma_k \right),
\end{equation}
where the operators $\sigma_{\pm}=\frac 12(\sigma_x \pm i \sigma_y)$ and the decay rates are also obtained by
\begin{align}
\label{eq:(7)}
\gamma_{-}&=2\lambda^2 \; \mathcal{G}(\omega_0), \nonumber \\
\gamma_{+}&=2\lambda^2 \;  \mathcal{G}(-\omega_0), \nonumber \\
\gamma_{0}&=2\lambda^2 \;  \mathcal{G}(0).
\end{align}
$\mathcal{G}(\omega)$ is determined by the Fourier transformation of the field vacuum correlation function $G^{+}$,
\begin{equation}
\label{eq:(8)}
\mathcal{G}(\omega)= \int_{-\infty}^{+\infty} \mathrm{d} s \; e^{i\omega s} G^{+}[x(s)].
\end{equation}
Here the correlation function between two different positions for the vacuum field can be calculated as
\begin{align}
\label{eq:(9)}
G^{+}\left[ x(\tau), x^{\prime}(\tau^{\prime})  \right]& = \langle 0_f \left| \phi\big(x(\tau) \big) \phi \big(x^{\prime}(\tau^{\prime}) \big) \right|0_f\rangle  \nonumber \\
&= -\frac {1}{4\pi^2}\; \dfrac {1}{(ct-ct^{\prime}-i\epsilon)^2-(x-x')^2}
\end{align}
with the proper $i\epsilon$ prescription. In this case, the initial position of the detector is chosen to be $\tau^{\prime}=0,x^{\prime}=0$.
In the quantum master equation, the effective Hamiltonian of the detector is expressed as $H_{eff}=\frac 12 \Omega \sigma_z$ where the effective energy gap
\begin{equation}
\label{eq:(10)}
\Omega= \omega_0+i\lambda^2[\mathcal{K}(-\omega_0)-\mathcal{K}(\omega_0)],
\end{equation}
where the parameter $\mathcal{K}(\omega)=\frac {\mathbf{P}}{\pi i}\int_{-\infty}^{+\infty} \mathrm{d} \omega^{\prime} \; \frac {\mathcal{G}(\omega^{\prime})}{\omega^{\prime}-\omega}$ and $\mathbf{P}$ denotes the principle value.

To mathematically express the density matrix $\rho(\tau)$, we equivalently replace the density matrix by the Bloch vector, i.e., $\rho=\dfrac {\mathbbm{1}+\sum_{j=x,y,z} B_j \; \sigma_j}{2}$. The components of the Bloch vector for the state are obtained by
\begin{equation}
\label{eq:(11)}
\frac {\mathrm{d} \overrightarrow{ \mathbf{B}}}{\mathrm{d}\tau} \;=\; U \overrightarrow{ \mathbf{B}} + \overrightarrow{ \mathbf{v}},
\end{equation}
where the unitary matrix $U$ for the dynamical map and inhomogeneity vector $\vec{\mathbf{v}}$ take the form
\begin{widetext}
\begin{equation}
\label{eq:(12)}
U =-\frac 12 \begin{pmatrix}
         \gamma_+ +\gamma_- +4\gamma_z& 2\Omega & 0\\
         -2\Omega & \gamma_+ +\gamma_- +4\gamma_z &0 \\
         0 & 0 & 2(\gamma_+ + \gamma_-)
         \end{pmatrix}, \; \; \overrightarrow{\mathbf{v}}= \begin{pmatrix}
                                                            0\\
                                                            0\\
                                                            \gamma_+-\gamma_-
                                                            \end{pmatrix}
\end{equation}
\end{widetext}
By means of the above equantion, we can conveniently attain the quantum evolution in the non-inertial referrence frame. The estimation information with respect to the phase parameter is embodied in the dynamical map.

\section{phase estimation and results}

Our aim is to evaluate relativistic effects on dynamical behavior of parameter quantum estimation for an accelerated atom. In experiments, the observation of a phase difference between two energy states plays a great role in quantum information technology \cite{Giovanetti04}. In this scenario, we consider phase estimation for atoms in relativistic motions. For a measuring device in the same motion as a uniformly accelerated atom, the observation of quantum estimation should be presented in the non-inertial frame. Different from quantum estimation in the non-inertial frame, the results are established in the inertial frame when the measuring device is in a motion with uniform velocity.

An selected initial state for the atom at $\tau=t=0$ involves a phase parameter and has the form of
\begin{equation}
\label{eq:(13)}
|\Psi_0(\varphi)\rangle= \cos\frac {\theta}{2}|1\rangle+\sin\frac {\theta}{2} \; e^{i\varphi}|0\rangle,
\end{equation}
where $\varphi$ represent a phase difference between the excited state $|1\rangle$ and ground state $|0\rangle$. $\theta$ describes the relative occurrence possibility of the energy states. It is easily found that the phase quantum estimation is dependent on the evolution of the parameterized states. To evaluate the true value of a phase parameter $\varphi$ as precisely as possible, we need an unbiased estimator $\boldsymbol{\hat{\varphi}}$ whose expectation holds that $\mathrm{Tr}(\rho_{\varphi} \, \boldsymbol{\hat{\varphi}})=\varphi$. The precision of quantum estimation satisfies the quantum Cram\'{e}r-Rao ($\mathrm{QCR}$) inequality \cite{Helstrom1976,Holevo1982}, i.e., $\Delta \boldsymbol{\hat{\varphi}}\ge \dfrac 1{\sqrt {M \; F(\rho_{\varphi})}}$, where $M$ is the number of independent measurements and $F(\rho_{\varphi})$ is the $\mathrm{QFI}$ with respect to phase parameter $\varphi$. The $\mathrm{QCR}$ inequality shows that the high precision of phase estimation is attained when the value of $\mathrm{QFI}$ is large. The standard calculation procedure \cite{Helstrom1976,Holevo1982} starts by the construction of a symmetric logarithmic derivative $L_{\varphi}$ which is defined as
\begin{equation}
\label{eq:(14)}
\frac {\partial}{\partial \varphi}\rho_{\varphi} = \partial_{\varphi} \rho_{\varphi} \;=\; \frac 12(L_{\varphi}\rho_{\varphi}+\rho_{\varphi}L_{\varphi}).
\end{equation}
The $\mathrm{QFI}$ which is not dependent on the choice of $L_{\varphi}$ is generally written as
\begin{equation}
\label{eq:(15)}
F(\rho_{\varphi})=\mathrm{Tr} [\rho_{\varphi} \, L^2_{\varphi}].
\end{equation}

The evolution state $\rho_{\varphi}(\tau)$ of the atom in the non-inertial frame is written as
\begin{align}
\label{eq:(16)}
B_x(\tau) &= e^{-\frac 12(\gamma_+ +\gamma_- +4\gamma_z)\tau} \; \sin \theta \; \cos(\Omega \tau+\varphi)\nonumber \\
B_y(\tau) &= e^{-\frac 12(\gamma_+ +\gamma_- +4\gamma_z)\tau} \; \sin \theta \; \sin(\Omega \tau+\varphi)\nonumber \\
B_z(\tau) &= e^{-(\gamma_+ +\gamma_- )\tau} \; \left(\cos\theta -\frac {\gamma_+ - \gamma_-}{\gamma_+ +\gamma_-} \right)+\frac {\gamma_+ - \gamma_-}{\gamma_+ +\gamma_-}
\end{align}
For the density matrix of $\rho_{\varphi}(\tau)=\sum_{j=1,2} p_{j}(\tau)|\psi_j(\tau)\rangle \langle \psi_j(\tau)|$, the elements of $L_{\varphi}$ are calculated as
\begin{equation}
\label{eq:(17)}
\left( L_{\varphi}\right)_{ij} = \frac 2{p_i+p_j}\left[ \sum_{i=1,2} (\partial_{\varphi}p_i) \, \delta_{ij}+p_j\langle \psi_i|\, (\partial_{\varphi}|\psi_j\rangle) + p_i(\partial_{\varphi}\langle \psi_i|)\, |\psi_j\rangle \right],
\end{equation}
where the eigenvalues and eigenstates of the density matrix are given by
\begin{align}
\label{eq:(18)}
p_{j} & = \frac {1\pm |\overrightarrow{B}|}{2} \nonumber \\
|\psi_j\rangle & = \dfrac 1{\sqrt{h^2+w^2_j}} \big(  \pm w_j \; e^{-i(\Omega \tau+\varphi)} |1\rangle \;+ \;h |0\rangle \big)
\end{align}
with the parameters of $h=e^{-\frac 12(\gamma_+ +\gamma_- +4\gamma_z)\tau} \sin \theta$ and $w_j=|\overrightarrow{B}|\pm B_z$. As a result, we accomplish the analytical expression of $\mathrm{QFI}$ as
\begin{equation}
\label{eq:(19)}
F_{\varphi}(\tau)= h^2.
\end{equation}
When the parameter $\theta=\pi/2$, the value of $\mathrm{QFI}$ reaches a maximal one, i.e., $F^{\mathrm{max}}_{\varphi}(\tau)=e^{(\gamma_+ +\gamma_- +4\gamma_z)\tau}$. It is seen that the dynamical behavior of $\mathrm{QFI}$ are determined by the decay rates and proper time.

To discover some controllable conditions that contribute to the $\mathrm{QFI}$, we reasonably take into account boundary effects of the scalar field. Boundaries can modify the fluctuations of quantum fields. It leads to a lot of novel effects, such as Casimir effects\cite{Casimir1948}, entanglement generation \cite{Zhang07} and the modification for the radiative properties of uniformly accelerated atoms \cite{Yu05}. In our scheme, two perfectly reflecting plane boundaries that are perpendicular to each other locate at $Y=Y_0$ and $Z=Z_0$ in the space, respectively. From Fig. (1), it is shown that $R$ denotes the distance between the trajectory of the accelerated atom and the cross of the two boundaries. The angle $\alpha$ represents the relative position of the atom between the two boundaries. In this simple case, the boundary conditions holds that $\phi|_{Y=Y_0}=0$ and $\phi|_{Z=Z_0}=0$. According to the method of images, the correlation function for the vacuum field is expressed as
\begin{align}
\label{eq:(20)}
G^{+} = -\frac {a^2}{16\pi^2} & \left\{ \dfrac 1{\sinh^2\left[ \frac {a(\tau-\tau^{\prime}}{2} -i\epsilon \right]} - \dfrac 1{\sinh^2\left[ \frac {a(\tau-\tau^{\prime}}{2} -i\epsilon \right]-a^2R^2\cos^2\alpha}  \right.  \nonumber \\
                                  & \left. + \dfrac 1{\sinh^2\left[ \frac {a(\tau-\tau^{\prime}}{2} -i\epsilon \right]-a^2R^2}  -  \dfrac 1{\sinh^2\left[ \frac {a(\tau-\tau^{\prime}}{2} -i\epsilon \right]-a^2R^2\sin^2\alpha} \right\}.
\end{align}
The Fourier transformation of the correlation function $\mathcal{G}(\omega)$ is also calculated as
\begin{equation}
\label{eq:(21)}
\mathcal{G}(\omega) = \frac {\omega}{2\pi(1-e^{-2\pi\omega/a})} \left[ 1-f^{(\omega)}_1(R\cos\alpha)-f^{(\omega)}_1(R\sin\alpha)+f^{(\omega)}_1(R) \right],
\end{equation}
where the special function $f_1$ is defined as $f^{(\omega)}_1(r)=\dfrac {\sin \left[ \frac {2\omega}{a}\; \sinh^{-1}(ar) \right]}{2r\sqrt{1+a^2r^2}\omega}$. The Unruh temperature $\frac {\omega}{2\pi}$ is involved in the above equation. And in the limit of $\omega \rightarrow 0$, the parameter $\mathcal{G}(0)=\frac {a}{4\pi^2}[1-f_2(R\cos\alpha)-f_2(R\sin\alpha)+f_2(R)]$ where the function $f_2(r)=\lim_{\omega \rightarrow 0} f^{\omega}_1(r)=\dfrac {\sinh^{-1}(ar)}{ar\sqrt{1+a^2r^2}}$. Consequently, the analytical expression of the maximal $\mathrm{QFI}$ in the non-inertial frame is given by
\begin{align}
\label{eq:(22)}
F^{\mathrm{max}}_{\varphi}(\tau)= \exp &  \left\{ -\dfrac {\lambda^2\omega_0\coth(\frac {\omega_0\pi}{a})\tau}{\pi}\left[ 1-f^{(\omega_0)}_1(R\cos\alpha)-f^{(\omega_0)}_1(R\sin\alpha)+f^{(\omega_0)}_1(R)  \right]  \right. \nonumber \\
                                           &- \left. \frac {\lambda^2 a \tau}{\pi^2}\left[ 1-f_2(R\cos\alpha)-f_2(R\sin\alpha)+f_2(R) \right] \right\}.
\end{align}

To clearly demonstrate the effects of the relativistic motions and boundaries on the dynamics of the $\mathrm{QFI}$, we carry out the numerical calculation of $\mathrm{QFI}$ which are plotted in Figs. (2)-(4). Figs. (2)-(4) provide the observation results of the maximal $\mathrm{QFI}$ in the non-inertial frame. The observation is achieved by a measuring device in the same motion as the acceleration atom. It is clearly seen that the values of the maximal $\mathrm{QFI}$ are always monotonically decreased with the proper time $\tau$ in Fig. (2). Besides it, the values in the low acceleration condition are larger than those in the high acceleration condition. This means that the relativistic motion of the measuring device can suppress the precision of parameter quantum estimation for the accelerated atom. The boundary effects on the $\mathrm{QFI}$ are shown in Fig. (3). In the region of $R\ll \frac 1{\omega_0}$ , the values of $\mathrm{QFI}$ can keep high in the short time interval. When the trajectory of the accelerated atom is far away from the boundaries, i.e., $R\gg \frac 1{\omega_0}$, the boundary effects are trivial and even neglected. We also notice that the apparent oscillation of the $\mathrm{QFI}$ will happen if the accelerated atom is close to the boundaries. This phenomena is induced by the relativistic motion. From Fig. (4), we also see that the values of $\mathrm{QFI}$ are symmetrical to the boundaries. In the case of $\alpha=\pi/4$, the value of $\mathrm{QFI}$ arrives at the minimal one. Meanwhile, the high values of $\mathrm{QFI}$ can be reached under the condition of $\alpha \rightarrow 0$ or $\alpha \rightarrow \pi/2$. The fact is that the precision of phase quantum estimation is enhanced when the accelerated atom is near to the boundaries.

\section{Discussion}

By means of the $\mathrm{UdW}$ model, we study the dynamical behavior of phase quantum estimation for a uniformly accelerated atom weakly coupled to a massless quantum scalar field. The dynamics of quantum states of the atom can be obtained in the non-inertial frame. The calculation of the $\mathrm{QFI}$ provides an efficient way to estimate the measuring precision. When a measuring device is accelerated in the same motion as the atom, the monotonic decrease of the $\mathrm{QFI}$ is manifest in the high acceleration limit. This means that the relativistic motion of the measuring device restrains the precision of phase quantum estimation. It is found out that the obvious oscillation of $\mathrm{QFI}$ occur under the condition that the trajectory of the atom is close to the boundaries. This provides us a possible way to enhance the precision of phase quantum estimation in the relativistic situation. Let us examine the magnitudes involved in some possible experiments. The natural scale of units is fixed by fixing units for the energy gap $\omega_0$ of the atom, namely $\tilde{a}=a(\omega_0 c)/\pi$. For atomic gaps of $\mathrm{GHz}$, one natural unit of acceleration corresponds to $10^{16}g$ ($g$ is the Earth surface gravitation acceleration) \cite{Martin13}. The acceleration required in our scenario can be reduced by the use of qubits with smaller gaps. For example, for special qubits defined by nuclear spins, energy gaps can be limited to the order of $\mathrm{MHz}$, which can reduce the acceleration to almost $10^{12}g$ \cite{Arimondo1977}. The scale of acceleration can be realized in $\mathrm{LHC}$ setups. In addition, qubits with Stark shifted atomic levels or Zenner-induced transition nearly have energy gaps of the order of $\mathrm{Hz}$. For these qubits, the acceleration can reach $10^{6}g$ \cite{Aad10}. Those accelerations are indeed experimentally achieved for short time intervals. Current technology of ion trapping and superconducting circuits allows for experiments where relativistic effects can be observed.

\begin{acknowledgments}
The work was supported by the $\mathrm{333}$ talents project of Jiangsu Province, by Jiangsu government scholarship for study abroad, and by NSERC of Canada and Templeton foundation, grant No. 36838. We would like to thank Bill Unruh for the discussions on related work.
\end{acknowledgments}

\newpage

{\Large \bf Figure Captions}

{\bf Fig. 1}

The accelerated atom is located between two reflecting plane boundaries perpendicular to each other. The method of images is used to calculate the correlation function for the vacuum field.

{\bf Fig. 2}

The effects of the acceleration on the dynamics of $\mathrm{QFI}$ as a function of the proper time $\tau$ are shown when the measuring device is in the same relativistic motion as the atom. The observation results are obtained under the conditions that $R=0.1, \; \alpha=0.1\pi, \; \omega_0=10, \; \lambda=1$ in the non-inertial frame.

{\bf Fig. 3}

The boundary effects on the dynamics of $\mathrm{QFI}$ as a function of the proper time $\tau$ are shown when the measuring device is in the same relativistic motion as the atom. The observation results are obtained under the conditions that $a=1, \; \alpha=0.1\pi, \; \omega_0=10, \; \lambda=1$ in the non-inertial frame.

{\bf Fig. 4}

The effects of the relative position of the atom between the boundaries are shown at some proper time when the measuring device is in the same relativistic motion as the atom. The observation results are obtained under the conditions that $\tau=0.4, \; a=1, \; R=0.4, \; \omega_0=10, \; \lambda=1$ in the non-inertial frame.

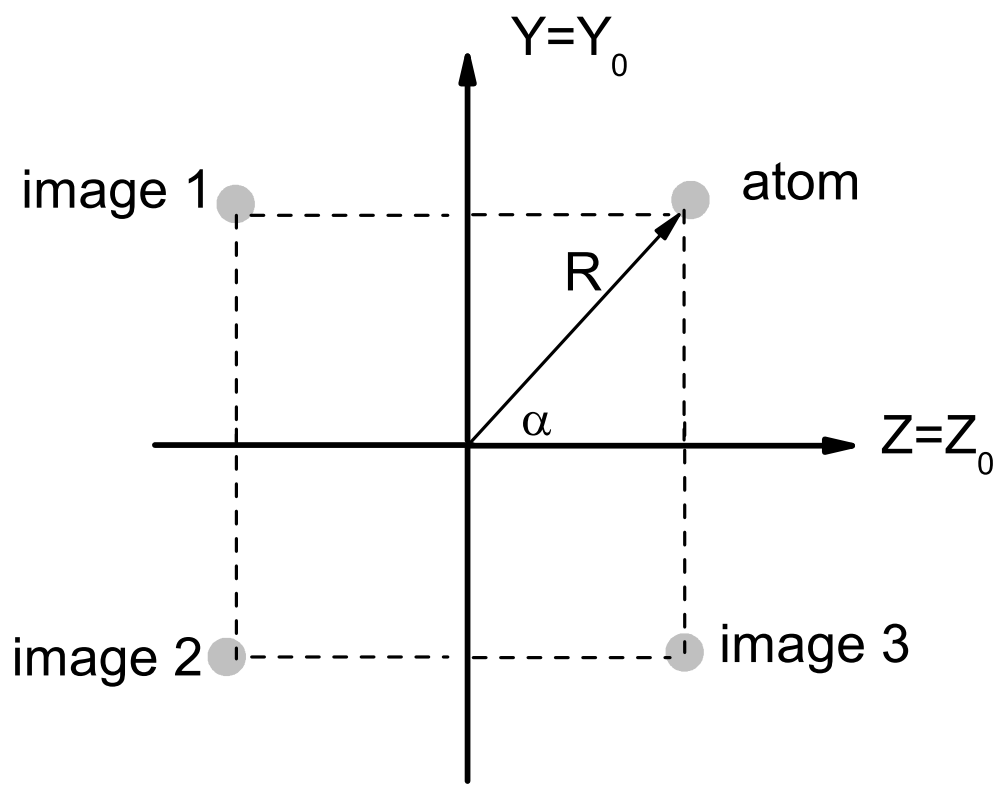
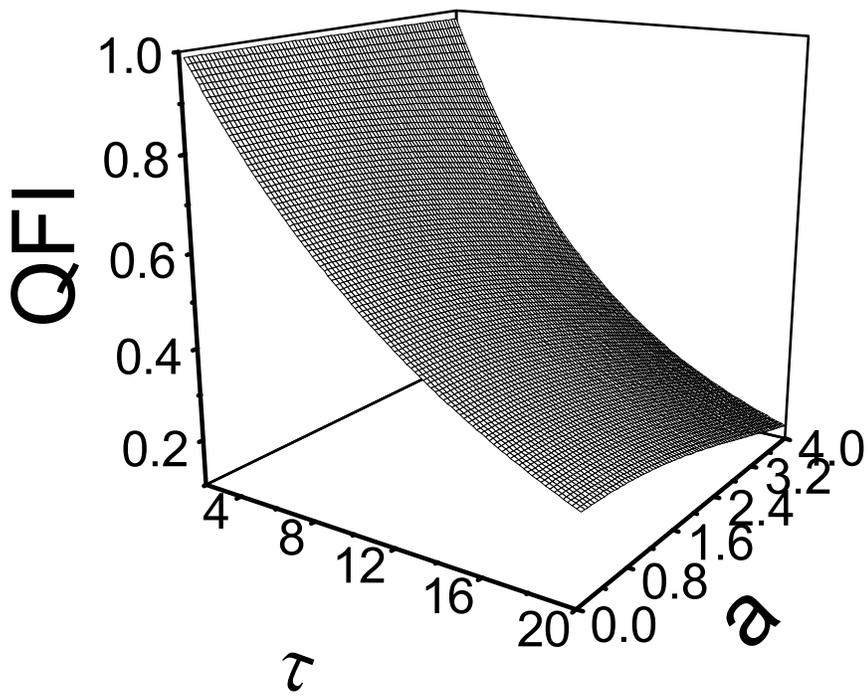
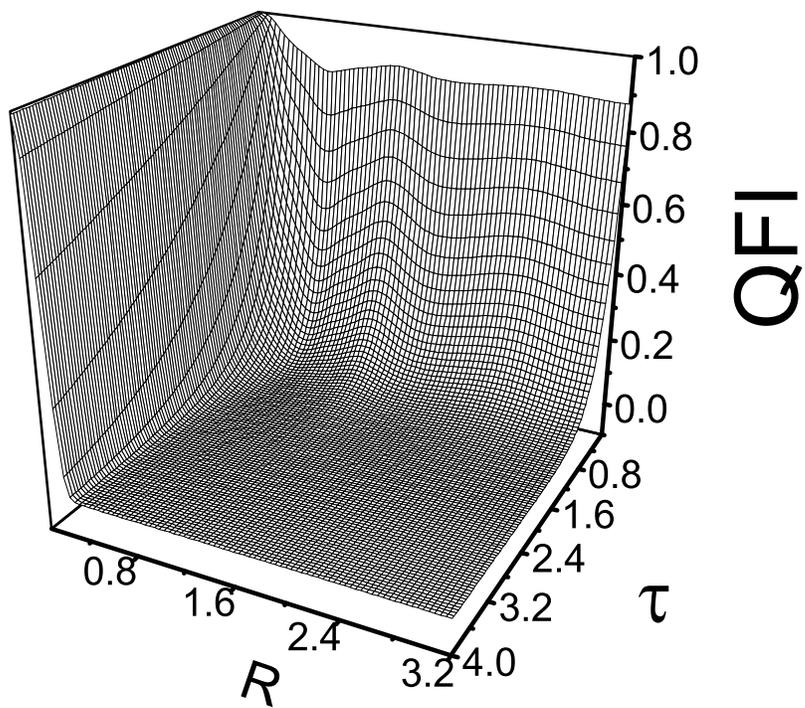
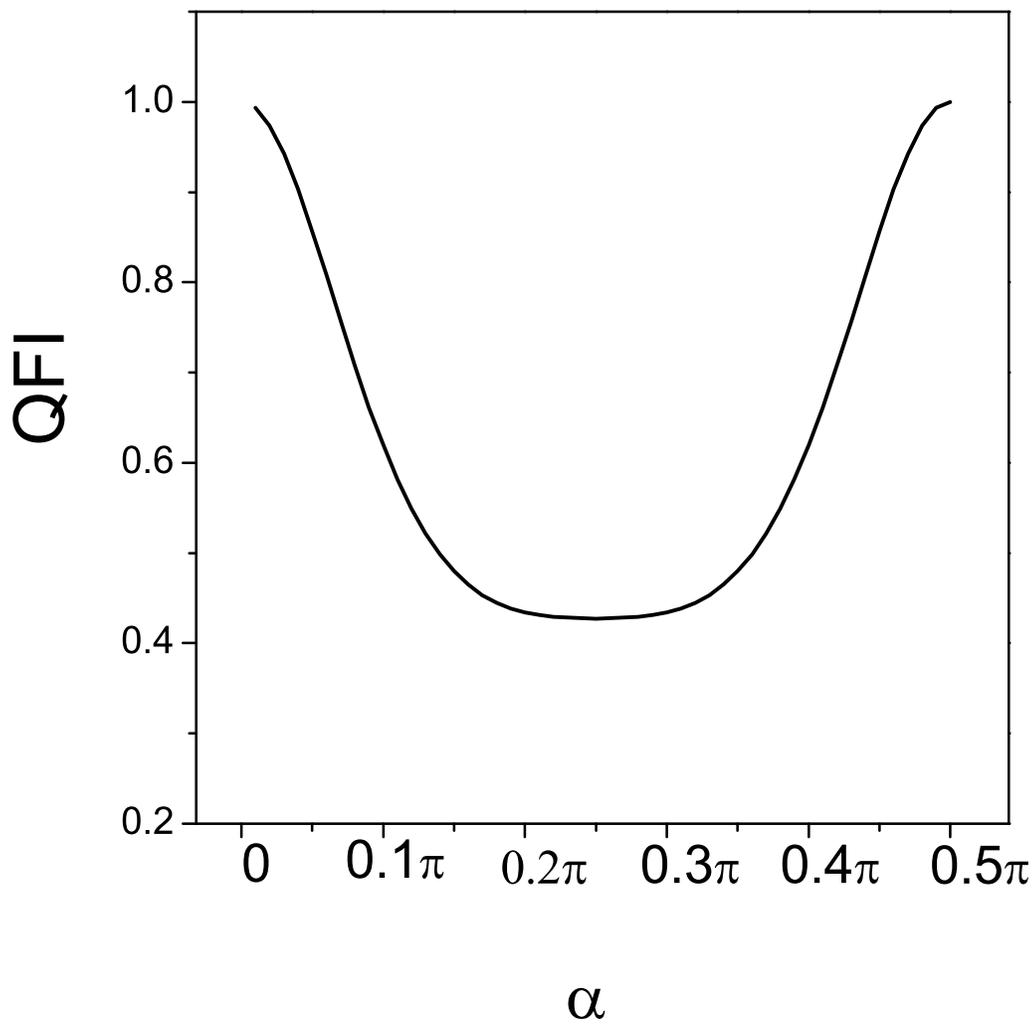
\end{document}